\documentclass[prd,showpacs,amsmath,showkeys,onecolumn,floatfix, nofootinbib]{revtex4} 

\usepackage{bm}
\usepackage{slashed}
\usepackage{graphicx}
\usepackage{subfigure}
\usepackage[svgnames]{xcolor}
\usepackage[normalem]{ulem}

\begin{document}

%\preprint{
\begin{flushright} ADP-10-21/T717 \end{flushright}
%}

\title{Calculating Kaon Fragmentation Functions from NJL-Jet Model}
\author{Hrayr H.~Matevosyan}
\author{Anthony W.~Thomas}
\affiliation{CSSM, School of Chemistry and Physics, \\
University of Adelaide, Adelaide SA 5005, Australia}
\author{Wolfgang Bentz}
\affiliation{Department of Physics, School of Science,\\  Tokai University, Hiratsuka-shi, Kanagawa 259-1292, Japan}

\begin{abstract}
The Nambu--Jona-Lasinio (NJL) - Jet model provides a sound framework for calculating the fragmentation functions in an effective chiral quark theory, where the  momentum and isospin sum rules are satisfied without the introduction of ad hoc parameters. Earlier studies of the pion fragmentation functions using the NJL model within this framework showed qualitative agreement with the empirical parameterizations. Here we extend the NJL-Jet model by including the strange quark. The corrections to the pion fragmentation functions and corresponding kaon fragmentation functions are calculated using the elementary quark to quark-meson fragmentation functions from NJL. The results for the kaon fragmentation functionsa exhibit a qualitative agreement with the empirical parameterizations, while the unfavored strange quark fragmentation to pions is shown to be of the same order of magnitude as the unfavored light quark's. The results of these studies are expected to provide important guidance for the analysis of a large variety of semi-inclusive data.
\end{abstract}

\pacs{13.60.Hb,~13.60.Le,~12.39.Ki  }
\keywords{Kaon fragmentation, NJL-Jet}
\maketitle

\section{Quark Fragmentation and NJL-Jet Model}
\label{NJL-JET} 

 Quark fragmentation functions have long been of interest for analyzing hard scattering reactions  \cite{Field:1977fa,Altarelli:1979kv,Collins:1981uw,Jaffe:1996zw,Ellis:1991qj,Barone:2001sp,Martin:2003sk}. New experimental  efforts to extract distribution and fragmentation functions from deep-inelastic lepton-nucleon, proton-proton scattering and $e^{+}e^{-}$ annihilation data \cite{Anselmino:2008jk, Anselmino:2008sga,Hirai:2007cx,deFlorian:2007aj} have generated a renewed interest in the subject \cite{Afanasev:2007qh, Kotzinian:2008fe}. The analysis of the transversity quark distribution functions \cite{Barone:2001sp,Ralston:1979ys} and a variety of other semi-inclusive processes \cite{Sivers:1989cc,Boer:2003cm} also critically depend on the knowledge of the fragmentation functions, both unpolarized and polarized  \cite{ Amrath:2005gv, Bacchetta:2007wc}.

 The NJL-Jet model of Ref.~\cite{Ito:2009zc} provides a self-consistent framework for calculations of both quark distributions and fragmentation functions in an effective chiral quark theory. The advantage of the model is that  there are no ad hoc parameters introduced to describe fragmentation functions. The isospin and momentum sum rules are satisfied naturally in the product ansatz. In this work we extend the NJL-Jet model by introducing the strange quark, thus allowing fragmentation to both pions and kaons. The inclusion of the new channel is shown to bring significant corrections to the previously calculated fragmentation functions to pions leading to a better agreement with empirical parametrizations.
   
%%%%%%%%%%%%%NEW SECTION%%%%%%%%%%%%%
\section{Including the Strange Quark}
\label{strange}

\subsection{Strange Quark Constituent Mass and Coupling}

Throughout this article we employ the $SU(3)$ NJL model with a straightforward generalization of the $SU(2)$ Lagrangian with no additional free parameters (see for e.g. Refs.~\cite{Kato:1993zw,Klimt:1989pm,Klevansky:1992qe} for detailed reviews of the NJL model) and we consider only the pseudoscalar mesons. Thus introducing the strange quark in the NJL-Jet model involves calculating the strange quark distribution and fragmentation functions in the NJL framework, requiring the knowledge of the quark-meson (strange-light to kaon) coupling constant and the strange quark constituent mass. We calculate the quark-meson coupling using the same approach as Ref.~\cite{Bentz:1999gx}. The strange quark mass is chosen to best fit the model calculated kaon mass to the experimental value.

 The quark-meson coupling constant is determined from the residue at the pole in the quark-antiquark t-matrix at the mass of the meson under consideration. This involves the derivative of the familiar quark-bubble graph:

\begin{equation}
\label{EQ_VACUUM_BUBBLE}
\Pi(p)=2N_{c}i\int \frac{d^{4}k}{(2\pi)^{4}} Tr[\gamma_{5}S_{1}(k)\gamma_{5}S_{2}(k-p)],
\end{equation}

\begin{equation}
\label{EQ_COUPLING_KQQ}
\frac{1}{g^{2}_{mqQ}}=- \left( \frac{\partial \Pi(p)}{\partial p^{2}}  \right)_{p^{2}=m^{2}_{m}}.
\end{equation}
 Here $\mathrm{Tr}$ denotes the Dirac-trace and the subscripts on the quark propagators denote quarks of different flavor - also indicated by $q$ and $Q$, where the meson of type $m$ under consideration has the flavor structure $m=q\overline{Q}$. We evaluate Eq.~(\ref{EQ_VACUUM_BUBBLE}) in light-cone (LC) coordinates\footnote{ We use the following LC convention for Lorentz 4-vectors $(a^+,a^-,\mathbf{a_\perp})$, $a^\pm=\frac{1}{\sqrt{2}}(a^0\pm a^3)$ and $\mathbf{a_\perp} = (a^1,a^2)$. }, noting:

\begin{eqnarray}
\label{EQ_VB_X}
\Pi(p) & = & \int_{-\infty}^{\infty}dx \Pi(p,x),\\
\Pi(p,x)& = &2N_{c}ip_{-}\int \frac{ dk_{+} d^{2}k_{\perp}}{(2\pi)^{4}} Tr[\gamma_{5}S_{1}(k)\gamma_{5}S_{2}(k-p)],\\
k_{-} &=& xp_{-}.
\end{eqnarray}

The $\Pi(p,x)$ is evaluated using the complex residue theorem and we obtain for $p_\perp=0$:

\begin{eqnarray}
\label{EQ_VB_XEVAL}
\Pi(p,x)& = & -2N_{c} \frac{\Theta(x) \Theta(1-x)}{x (1-x)} \int \frac{ d^{2}k_{\perp}}{(2\pi)^{3}} \frac{ k_{\perp}^{2} + ((1-x)M_{1}+xM_{2})^{2} }{k_{\perp}^{2} + (1-x)M_{1}^{2} +xM_{2}^{2} -x(1-x)p^{2} -i\epsilon }.
\end{eqnarray}

 Thus for the quark-meson coupling we obtain:

\begin{eqnarray}
\label{EQ_MQQ_COUPL}
g_{mqQ}^{-2}& = & 2N_{c} \int_{0}^{1} dx \int \frac{ d^{2}k_{\perp}}{(2\pi)^{3}} \frac{ k_{\perp}^{2} + ((1-x)M_{1}+xM_{2})^{2} } {(k_{\perp}^{2} + (1-x)M_{1}^{2} +xM_{2}^{2} -x(1-x)m_{m}^{2} -i\epsilon)^{2} }.
\end{eqnarray}

  The integrals in the above expressions are divergent and require regularization. Here we use the  Lepage-Brodsky (LB) ``invariant mass'' cut-off regularization (see Refs.~\cite{Bentz:1999gx,Ito:2009zc} for a detailed description as applied to the NJL-Jet model), where $\Lambda_{12}$ is the maximum invariant mass of the two particles in the loop:

\begin{eqnarray}
\label{EQ_LB_REG}
M_{12}\leq \Lambda_{12} \equiv \sqrt{\Lambda_{3}^{2}+M_{1}^{2}} +\sqrt{\Lambda_{3}^{2}+M_{2}^{2}},\\
M_{12}^{2} = \frac{M_{1}^{2}+k_{\perp}^{2}}{x} + \frac{M_{2}^{2}+k_{\perp}^{2}}{1-x},\\
k_{\perp}^{2} \leq x(1-x)\Lambda_{12}^{2}-(1-x)M_{1}^{2}-xM_{2}^{2},
\end{eqnarray}
where $\Lambda_3$ denotes the 3-momentum cutoff, which is fixed in the usual way by reproducing the experimental pion decay constant. We use the value obtained in the Ref.~\cite{Bentz:1999gx} of $\Lambda_3= 0.67 ~{\rm GeV}$.

 Thus from the requirement of $k_{\perp}^{2}\geq 0$ we obtain the region of $x$ where the integrand is non-zero.
 
Using the light constituent quark mass $M=0.3~{\rm GeV}$ from the Ref.~\cite{Bentz:1999gx}  and the experimental value of kaon mass $m_{K}=0.495~{\rm GeV}$ yields a strange constituent quark mass $M_{s}=0.537~{\rm GeV}$ and the corresponding quark-kaon coupling constant of $g_{KqQ}=3.39$.

 \subsection{Quark Distribution and Fragmentation Functions}

The quark distribution function $f_{q}^{h}(x)$ has an interpretation as the probability to find a quark of type $q$ with momentum fraction $x$ in the hadron (in our case meson) $h$.  The corresponding cut diagram is shown in Fig.~\ref{PLOT_Q_DISTR}a, which can be equivalently represented by the Feynman diagram depicted in Fig.~\ref{PLOT_Q_DISTR}b:

\begin{figure*}[ptb]
\centering 
\includegraphics[width=0.9\textwidth]{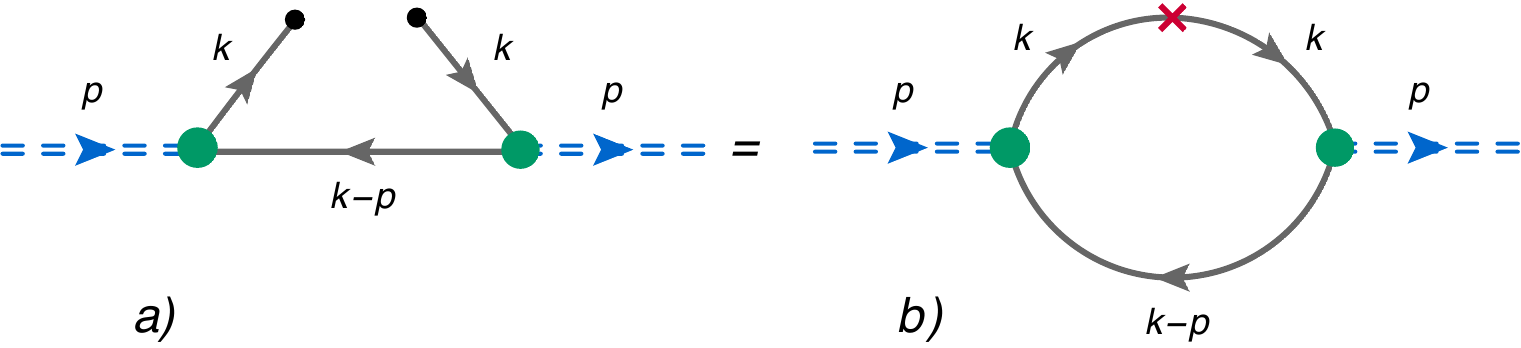}
\caption{Quark distribution functions.}
\label{PLOT_Q_DISTR}
\end{figure*}

\begin{eqnarray}
\label{EQ_Q_DISTR}
f_{q}^{m}(x)& = &i N_{c} \frac{C_q^m}{2}   g_{mqQ}^{2} \int \frac{ dk_{+} d^{2}k_{\perp}}{(2\pi)^{4}} Tr[\gamma_{5}S_{1}(k)\gamma^{+}S_{1}(k)\gamma_{5}S_{2}(k-p)] \\ \nonumber
&=& - g_{mqQ}^2 \frac{\partial \Pi(p,x)}{\partial p^2}|_{p^2=m_m^2} 
=N_{c} C_q^m g_{mqQ}^{2} \int \frac{ d^{2}k_{\perp}}{(2\pi)^{3}} \frac{k_{\perp}^{2}+((1-x)M_{1}+xM_{2})^{2}} {(k_{\perp}^{2}+(1-x)M_{1}^{2}+xM_{2}^{2}-x(1-x)m_{m}^{2})^{2}},
\end{eqnarray}
where $k_{-}= xp_{-}$ and $C_q^m$ is the corresponding flavor factor given in the Table~\ref{TB_FLAVOR_FACTORS}. Here we used the identity $S(k) \gamma^+ S(k) =-\partial S(k) \partial k_+$ to relate the distribution function to the function $\Pi(p,x)$ and the expression in the Eq.~(\ref{EQ_VB_XEVAL}) to evaluate the result.

\begin{figure}[ptb]
\centering 
\subfigure[] {
\includegraphics[width=0.48\textwidth]{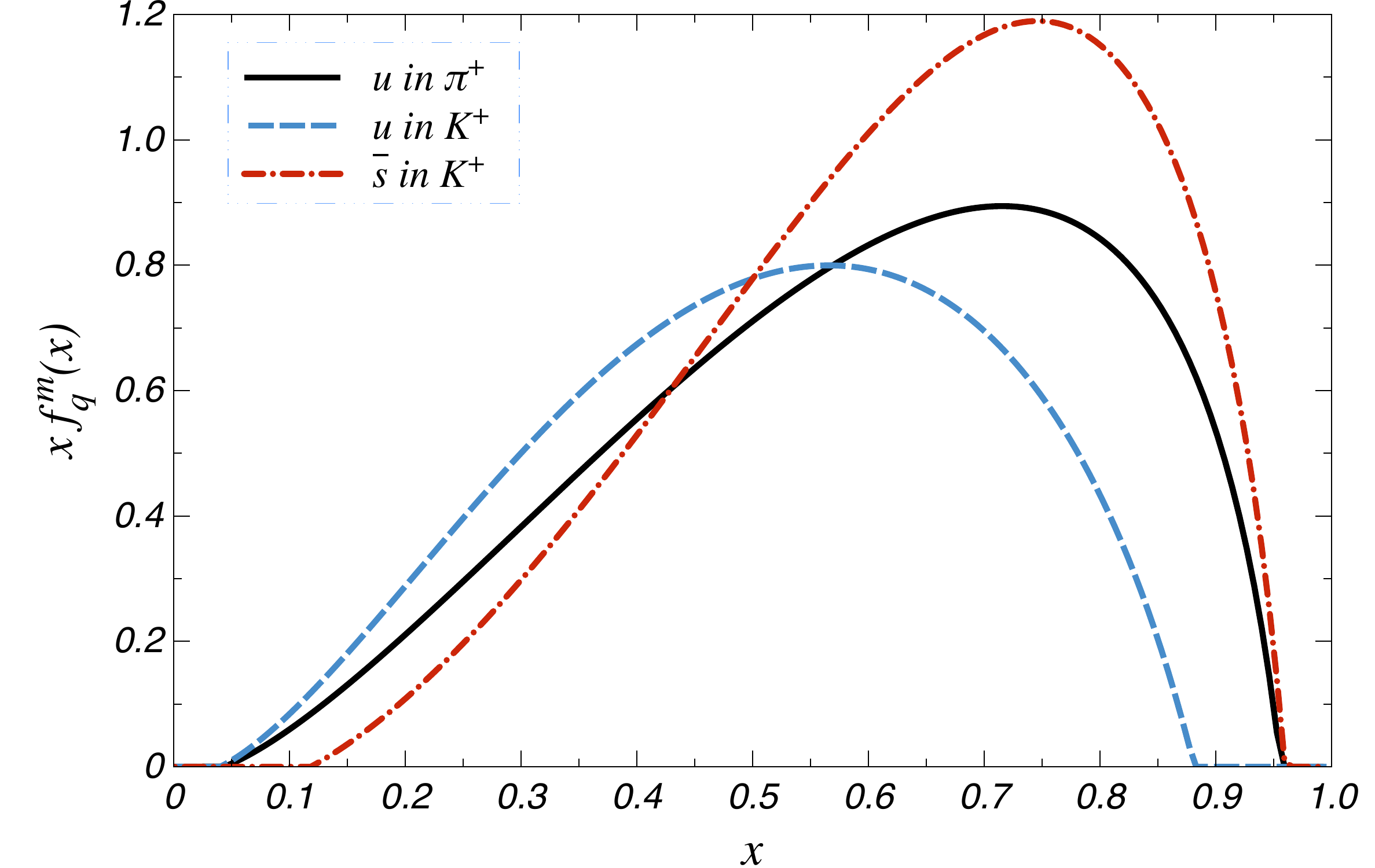}}
\hspace{0.1cm} 
\subfigure[] {
\includegraphics[width=0.48\textwidth]{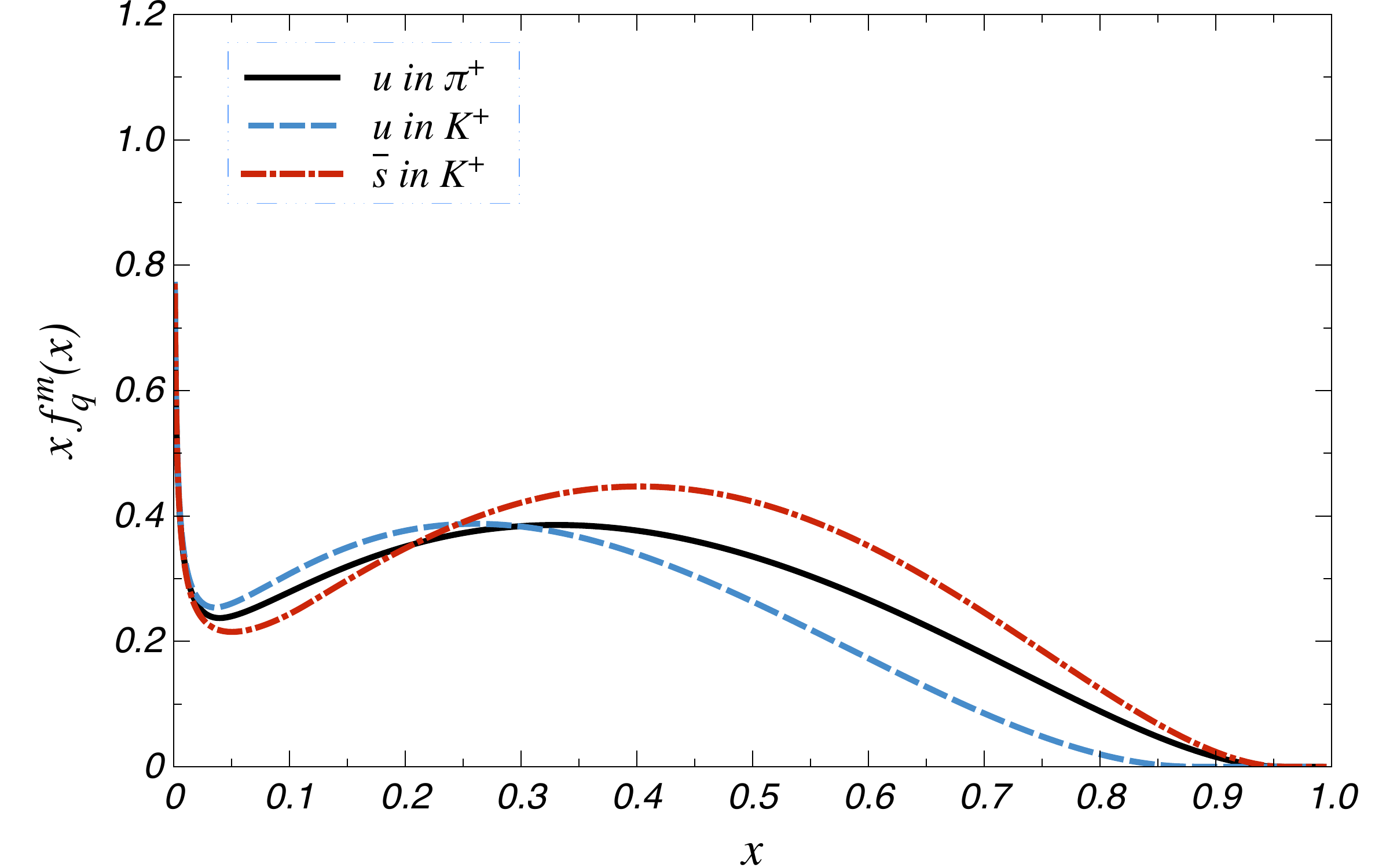}
}
\caption{The model solutions for distribution functions of quarks in mesons at a) model scale $Q_{0}^{2}=0.2~{\rm GeV}^{2}$ and b) evolved at next to leading order to scale $Q^{2}=4~{\rm GeV}^{2}$.}
\label{PLOT_Q_DISTR_CALCS}
\end{figure}

In our calculations we use the same LB regularization scheme (\ref{EQ_LB_REG}) yielding the non-zero region for $f_{q}^{m}(x)$ as described in the previous section. The corresponding plots for $f_{u}^{\pi^{+}}$, $f_{u}^{K^{+}}$ and $f_{\bar{s}}^{K^{+}}$  are depicted in Fig.~\ref{PLOT_Q_DISTR_CALCS} at model scale $Q_{0}^{2}=0.2~{\rm GeV}^{2}$ and evolved at next to leading order (NLO) to scale $Q^{2}=4~{\rm GeV}^{2}$ using the software from Ref.~\cite{Botje:2010ay}. Here the model scale is set by reproducing the behavior, after NLO evolution, of the u quark valence distribution function in the pion given in Refs.~\cite{Sutton:1991ay, Wijesooriya:2005ir}.

It is easy to check using Eqs.~(\ref{EQ_VB_X}, \ref{EQ_Q_DISTR}) that both isospin and momentum sum rules for the distribution functions are satisfied:

\begin{eqnarray}
\label{EQ_DISTR_SUMRULES}
\int_{0}^{1} dx \  f_{q}^{m}(x)& = &\frac{1}{2}C_q^m, \\
\int_{0}^{1} dx \ x\  \left(f_{q}^{m}(x) +f_{\overline{Q}}^{m}(x) \right) & = & \frac{1}{2}C_q^m ,\ m= q \overline{Q},
\end{eqnarray}
where in the last line we used the relations $f_q^m(x)=f_{\overline{Q}}^m(1-x)$ (which is easy to verify from Eq.~(\ref{EQ_Q_DISTR}) ) and $C_q^m=C_{\overline{Q}}^m$ (see the Table~\ref{TB_FLAVOR_FACTORS}).

\begin{figure}[ptb]
\centering 
\includegraphics[width=0.45\textwidth]{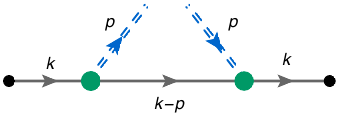}
\caption{Quark fragmentation functions.}
\label{PLOT_FRAG_QUARK}
\end{figure}

The elementary fragmentation function depicted in Fig.~\ref{PLOT_FRAG_QUARK} in the frame where the fragmenting quark has  $\textbf{k}_\perp=0$ (but non-zero transverse momentum component $k_T=-p_\perp/z$ with respect to the direction of the produced hadron) can be written as:

\begin{eqnarray}
\label{EQ_QUARK_FRAG}
\nonumber
d_{q}^{m}(z)&=& - \frac{C_q^m}{2}  g_{mqQ}^{2} \frac{z}{2} \int \frac{d^{4}k}{(2\pi)^{4}} Tr[S_{1}(k)\gamma^{+}S_{1}(k)\gamma_{5} (\slashed{k}-\slashed{p}+M_{2}) \gamma_{5}]\\ 
&& \times \delta(k_{-} - p_{-}/z) 2 \pi \delta( (p-k)^{2} -M_{2}^{2} )
= - \frac{z}{2N_{c}} f_{q}^{m}(x = 1/z)\\ 
&=& \frac{C_q^m}{2}  g_{mqQ}^{2} z \int \frac{ d^{2}p_{\perp}}{(2\pi)^{3}} \frac{p_{\perp}^{2}+((z-1)M_{1}+M_{2})^{2}} {(p_{\perp}^{2}+z(z-1)M_{1}^{2}+zM_{2}^{2}+(1-z)m_{m}^{2})^{2}}.
\end{eqnarray}

 The integration can be trivially done analytically if one assumes a sharp cut-off in the transverse momentum, denoted by $P^2_{\perp}$ ( in the LB regularization scheme $P^2_\perp$ is $z$-dependent):

\begin{eqnarray}
\label{EQ_QUARK_FRAG_ANAL}
d_{q}^{m}(z)&=& \frac{C_q^m}{2}  \frac{ g_{mqQ}^{2} }{ 8 \pi^2 } z \left( \frac{A/B-1}{B/P^{2}_{\perp} +1} + \log\left(1+P^{2}_{\perp}/B\right) \right), 
\end{eqnarray}

where

\begin{eqnarray}
\label{EQ_QUARK_FRAG_ANAL_DEFS}
A&\equiv& ((z-1)M_{1}+M_{2})^{2}, \\
B&\equiv& z(z-1)M_{1}^{2}+zM_{2}^{2}+(1-z)m_{m}^{2}.
 \end{eqnarray}

In LB regularization the $P_\perp^2$ is given by:

\begin{eqnarray}
\label{EQ_QUARK_FRAG_LB_PPERP}
P_\perp^2= z(1-z)\left(\sqrt{\Lambda_{3}^2 +m_m^2}+  \sqrt{\Lambda_{3}^2 +M_2^2}\right)^2 - (1-z)m_m^2 -z M_2^2.
 \end{eqnarray}

The corresponding sum rule for the elementary fragmentation function is as follows:
 \begin{eqnarray}
\label{EQ_ELM_FRAG_SUMRULES}
\int_{0}^{1} dz \  d_{q}^{m}(z)& =1-Z_{q}^m.
\end{eqnarray}

\begin{figure}[ptb]
\centering 
\includegraphics[width=0.45\textwidth]{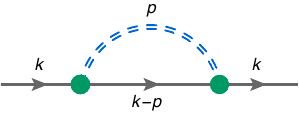}
\caption{(Color Online) Quark Self-Energy.}
\label{PLOT_QUARK_SE}
\end{figure}

Here $Z_{q}^m$ is the residue of the quark propagator of flavor $q$ in the presence of the cloud of the meson type $m$. It is expressed in terms of the renormalized quark self-energy induced by the meson type $m$ loop  $\Sigma_{q}^{m}(k)$ of Fig.~\ref{PLOT_QUARK_SE} as

\begin{eqnarray}
\label{EQ_Q_SE}
&1-Z_{q}^m=-\left(\frac{\partial \Sigma_{q}^{m}(k)}{\partial \slashed{k}} \right)_{\slashed{k}=M_q},\\
&\Sigma_{q}^{m}(k)= - \imath\ C_q^m g_{mqQ}^2\int \frac{d^4p}{(2\pi)^4} \gamma_5 S_Q(k-p) \gamma_5 \Delta_m(p),
\end{eqnarray} 
where $M_q$ is the mass of the constituent quark of flavor $q$ and $\Delta_m(p)$ is the Feynman propagator of the meson type $m$.

%%%%%%%%%%%%%NEW SECTION%%%%%%%%%%%%%
\section{Generalized NJL-Jet}
\label{SEC_NJL_JET}

The NJL-Jet model of Ref.~\cite{Ito:2009zc} uses a multiplicative ansatz for the total fragmentation function to derive an integral equation for the quark cascade for the process depicted in Fig.~\ref{PLOT_QUARK_CASCADE}. The derived integral equation for the total fragmentation function is:

\begin{figure*}[ptb]
\centering 
\includegraphics[width=0.65\textwidth]{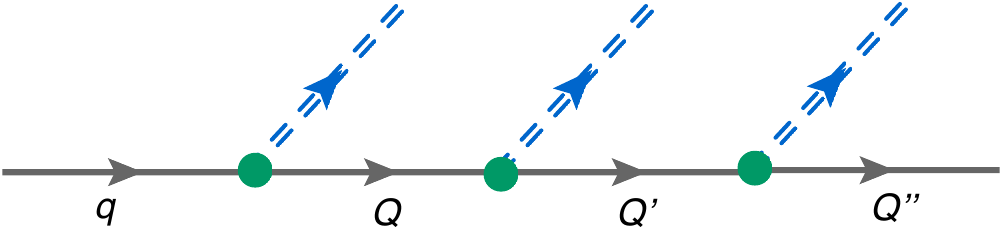}
\caption{Quark cascade.}
\label{PLOT_QUARK_CASCADE}
\end{figure*}

\begin{eqnarray}
\label{EQ_JET_INT}
D^{m}_{q}(z)=\hat{d}^{m}_{q}(z)+\sum_{Q}\int^{1}_{z}\frac{dy}{y}\hat{d}^{Q}_{q}(\frac{z}{y})\; D^{m}_{Q}(y), \hspace{20pt}
\hat{d}^{Q}_{q}(z)=\hat{d}^{m}_{q}(1-z)|_{m=q\bar{Q}} .
\end{eqnarray}

Here $\hat{d}_{q}^{m}(z) = d_{q}^{m}(z)/\sum_{m'}(1-Z_{q}^{m'})$ (here the sum is over all possible meson tpyes $m'$ that the quark of flavor $q$ can emit in the elementary splitting process).  Then $\sum_{m}\int \hat{d}_{q}^{m}(z) dz=1$, thus allowing an interpretation as the probability of an elementary process. In Eq.~(\ref{EQ_JET_INT}) the sum is over the flavor of the emitted quark $Q$ and the splitting function of quark $q$ into $Q$ with a light-cone momentum fraction $z$ is naturally the same as the splitting function of $q$ into a meson $m$ of a flavor composition $q\bar{Q}$ with a light-cone momentum fraction $1-z$.

 The result in Eq.(\ref{EQ_JET_INT}) resembles the integral equation ansatz of Field and Feynman's quark-jet model \cite{Field:1976ve,Field:1977fa}. Rewriting the above expression helps to elucidate the probabilistic interpretation of the model:

\begin{equation}
\label{ EQ_FRAG_PROB}
D^{m}_{q}(z)dz=\hat{d}^{m}_{q}(z)dz+\sum_{Q}\int^{1}_{z}\hat{d}^{Q}_{q}(y) dy  \; D^{m}_{Q}(\frac{z}{y}) \frac{dz}{y}.
\end{equation}

Here the left hand side term has the meaning of the probability to create a meson $m$ carrying the light-cone momentum fraction $z$ to $z+dz$ of initial quark $q$. The first term on the right hand side corresponds to the probability of creating the meson with light-cone momentum fraction $z$ to $z + dz$ in the first step of the cascade, plus the second term corresponding to the creation of the meson further down the quark cascade after a splitting to a quark $Q$ with light-cone momentum fraction $y$. Here the probability of creating the meson $m$ with the light-cone momentum fraction $z$ of the initial quark is the probability of creating the same meson with the light-cone momentum fraction $z/y$ in the cascade of the  quark $Q$, which is clearly only the case in the Bjorken limit. Thus the model can be generalized in a straightforward way by including the strange quark directly in Eq.~(\ref{EQ_JET_INT}). 

 We solve the coupled set of integral equations  using the elementary fragmentation functions of Eq.~(\ref{EQ_JET_INT}) for $u$, $d$ and $s$ quark fragmentation to a given meson. The corresponding fragmentation functions for the anti-quarks are obtained using charge symmetry starting from the fragmentation functions of quarks to the corresponding anti-meson. The comparisons with the phenomenological parametrizations of Ref.~\cite{Hirai:2007cx} are performed by DGLAP evolution of the calculated fragmentation functions from the low-energy model scale of $Q_0^{2}=0.2~{\rm GeV}^{2}$ to $4~{\rm GeV}^{2}$, at NLO, using the software from Ref.~\cite{Botje:2010ay}.

 It is easy to see, using the properties of $d_{q}^{m}$ along  with the normalization condition of $\hat{d}_{q}^{m}$, that the solutions of the integral equations (\ref{EQ_JET_INT}) should satisfy both momentum and isospin sum rules. We use a notation:
  
 \begin{equation}
\label{ EQ_MEAN_DEAF}
\left< f(z) \right> \equiv \int_0^1 f(z) dz,
\end{equation}

and define:

\begin{eqnarray}
\label{ EQ_MOM_SUMRULE}
N \equiv  \sum_m\left< z D_q^m(z) \right>, \ 
n \equiv  \sum_m\left< z \hat{d}_q^m(z) \right>, \ 
n' \equiv  \sum_{Q}\left< z \hat{d}_q^Q(z) \right>. 
\end{eqnarray}

 Using the equations Eq.~(\ref{EQ_JET_INT}) it follows:
 \begin{equation}
 N=n+n' N.
  \end{equation}
  
To prove the momentum sum rule $N  =1$, that is the fragmenting quark transfers all of its momentum to the emitted mesons, it is sufficient to show that $ n+ n '  =1$:

 \begin{eqnarray}
\label{ EQ_MOM_SUMRULE_PROVE}
n +n' =\sum_m \left( \left< z \hat{d}_q^m(z) \right>  + \left< z \hat{d}_q^m(1-z) \right>  \right) = \sum_m\left< \hat{d}_q^m(z) \right> =1 .
\end{eqnarray}

\begin{figure}[ptb]
\centering 
\subfigure[] {
\includegraphics[width=0.48\textwidth]{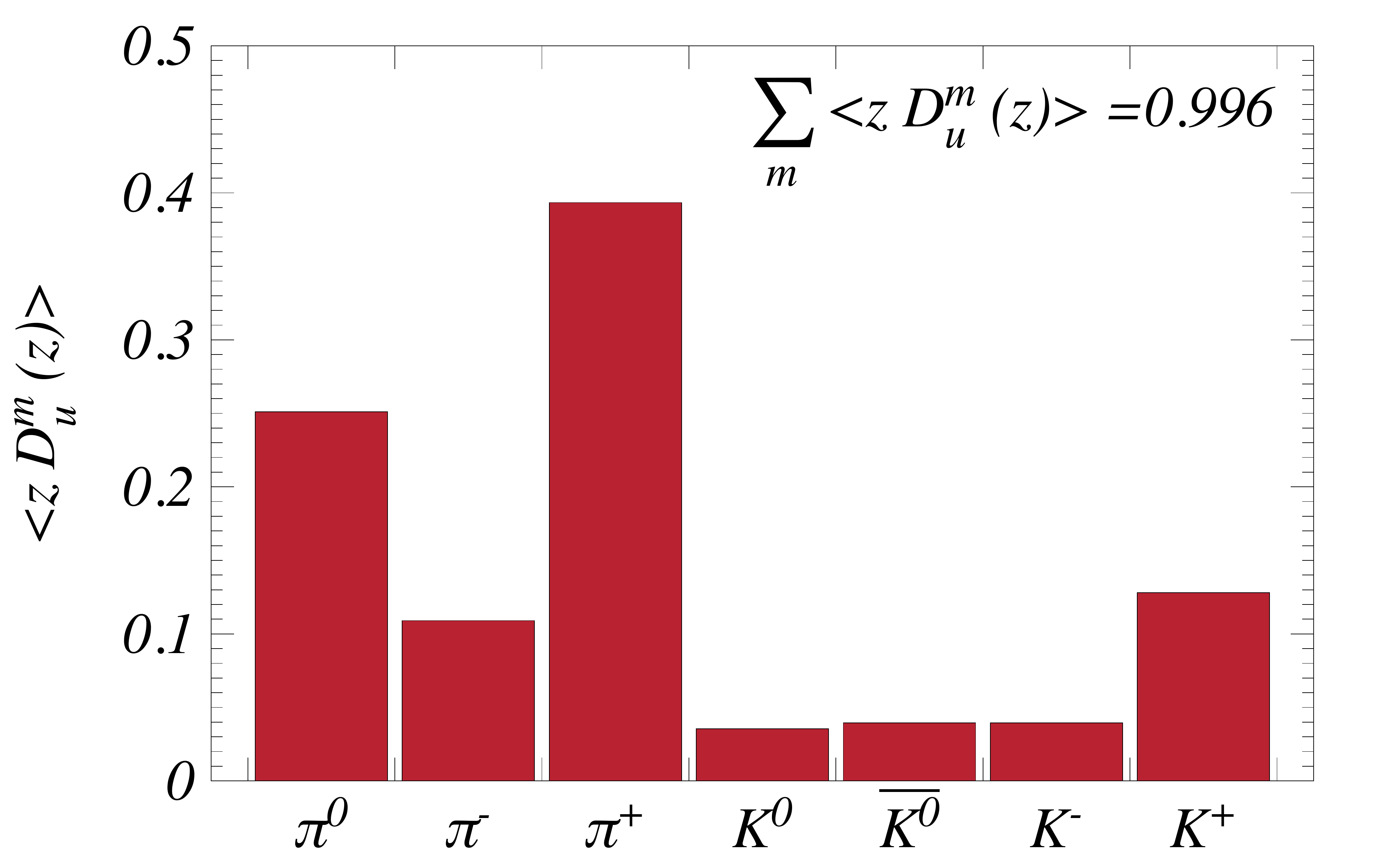}}
\hspace{0.1cm} 
\subfigure[] {
\includegraphics[width=0.48\textwidth]{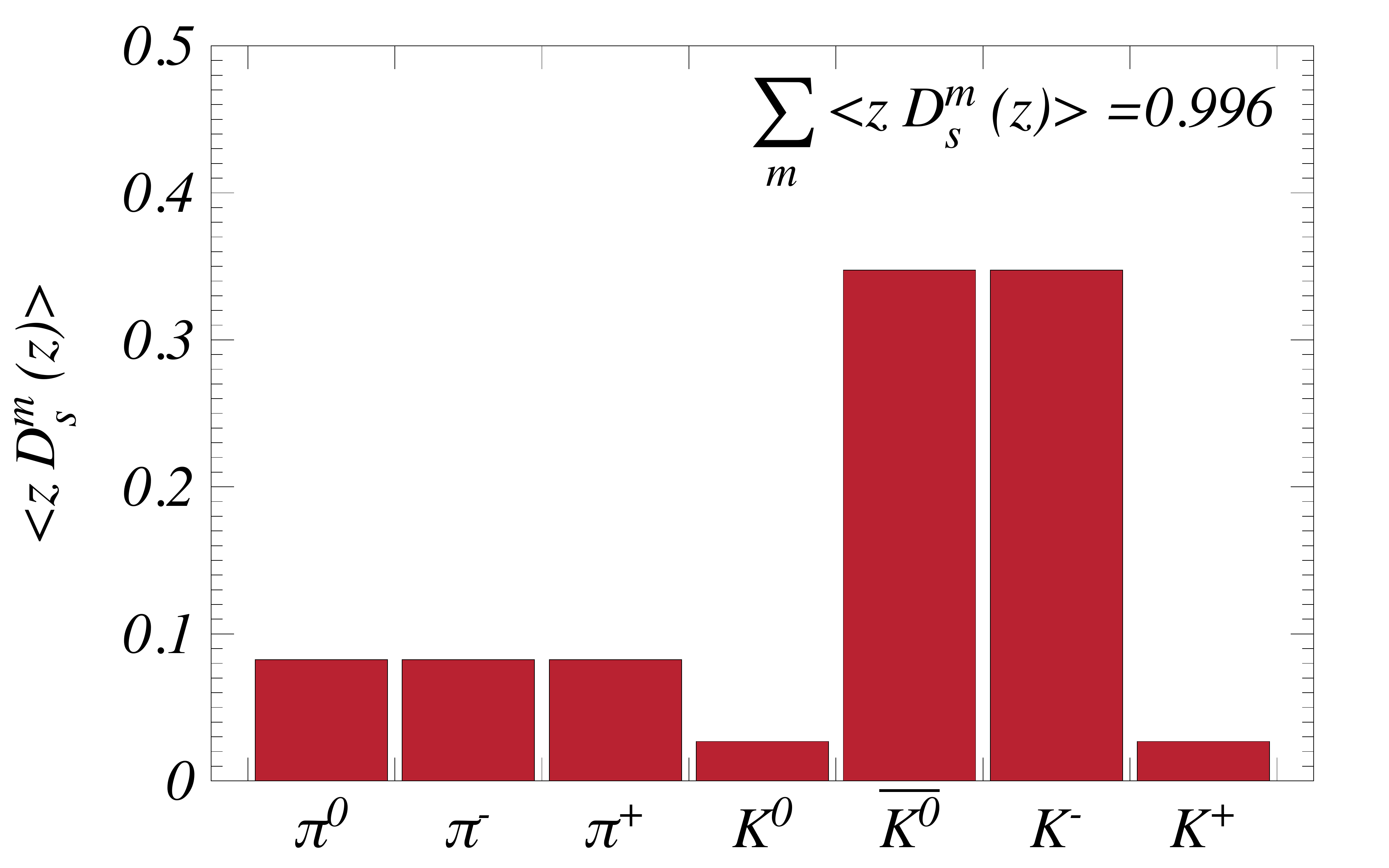}
}
\caption{The total fractions of momenta carried by mesons of type $m$ from the jet of  a) $u$ and b) $s$ quarks calculated using the numerical solutions for fragmentation functions.}
\label{PLOT_MOM_SUM}
\end{figure}

 The total fractions of momenta carried by mesons of type $m$ from the jet of $u$ and $s$ quarks, $\left< z D_q^m(z) \right>;\ q=u, s$ calculated using the numerical solutions for fragmentation functions are shown in Fig.~\ref{PLOT_MOM_SUM}, which show that the momentum sum rule is satisfied within the numerical precision of the calculation of less than a percent.
 
 Similarly we define $t$ as the z-component of the isospin and

\begin{eqnarray}
\label{EQ_IS_SUMRULE}
At_q\equiv \sum_m t_m \left< D_q^m(z)\right>,\ 
at_q\equiv \sum_m t_m \left< \hat{d}_q^m(z)\right>, \ 
a't_q\equiv \sum_Q t_Q \left< \hat{d}_q^Q(z)\right>.
\end{eqnarray}

Using the equations Eq.~(\ref{EQ_JET_INT}) it follows:
 \begin{equation}
 A=a+a' A.
  \end{equation}

To prove the isospin sum rule $A =1$, that is the z-component of the isospin is conserved in the quark fragmentation cascade, it is sufficient to show that $ a+ a '  =1$: 

 \begin{eqnarray}
\label{ EQ_IS_SUMRULE_PROVE}
at_q + a' t_q =\sum_{\begin{array}{c}m \\ m=q\bar{Q} \end{array} } \left( \left< t_h \hat{d}_q^m(z) \right>  + \left< t_Q \hat{d}_q^m(1-z) \right>  \right)  = t_q\sum_m\left< \hat{d}_q^m(z) \right> =t_q ,
\end{eqnarray}
where we kept the flavor of the fragmenting quark $q$ constant and the sum is over all possible mesons $m$ that can be emitted (for each $m$ the corresponding fragmented quark is denoted $Q$). 

Our numerical solutions obey these rules within  numerical errors of less than a percent, with, for example,  $86\%$ of the $u$ quark's isospin transferred to pions and $14\%$ to kaons. 

\section{Results and Conclusions}
\label{results}

\begin{figure}[ptb]
\centering 
\subfigure[] {
\includegraphics[width=0.48\textwidth]{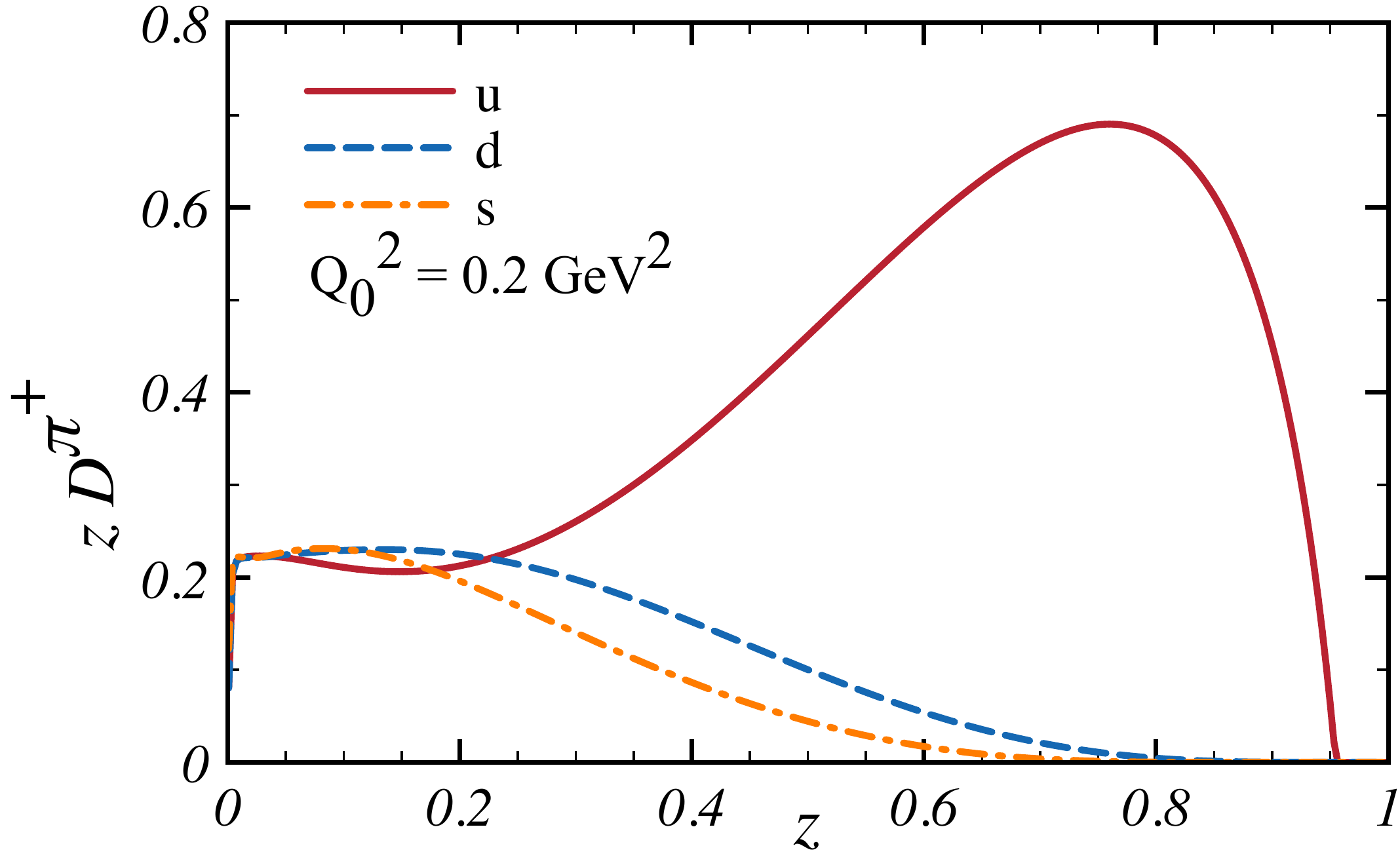}}
\hspace{0.1cm} 
\subfigure[] {
\includegraphics[width=0.48\textwidth]{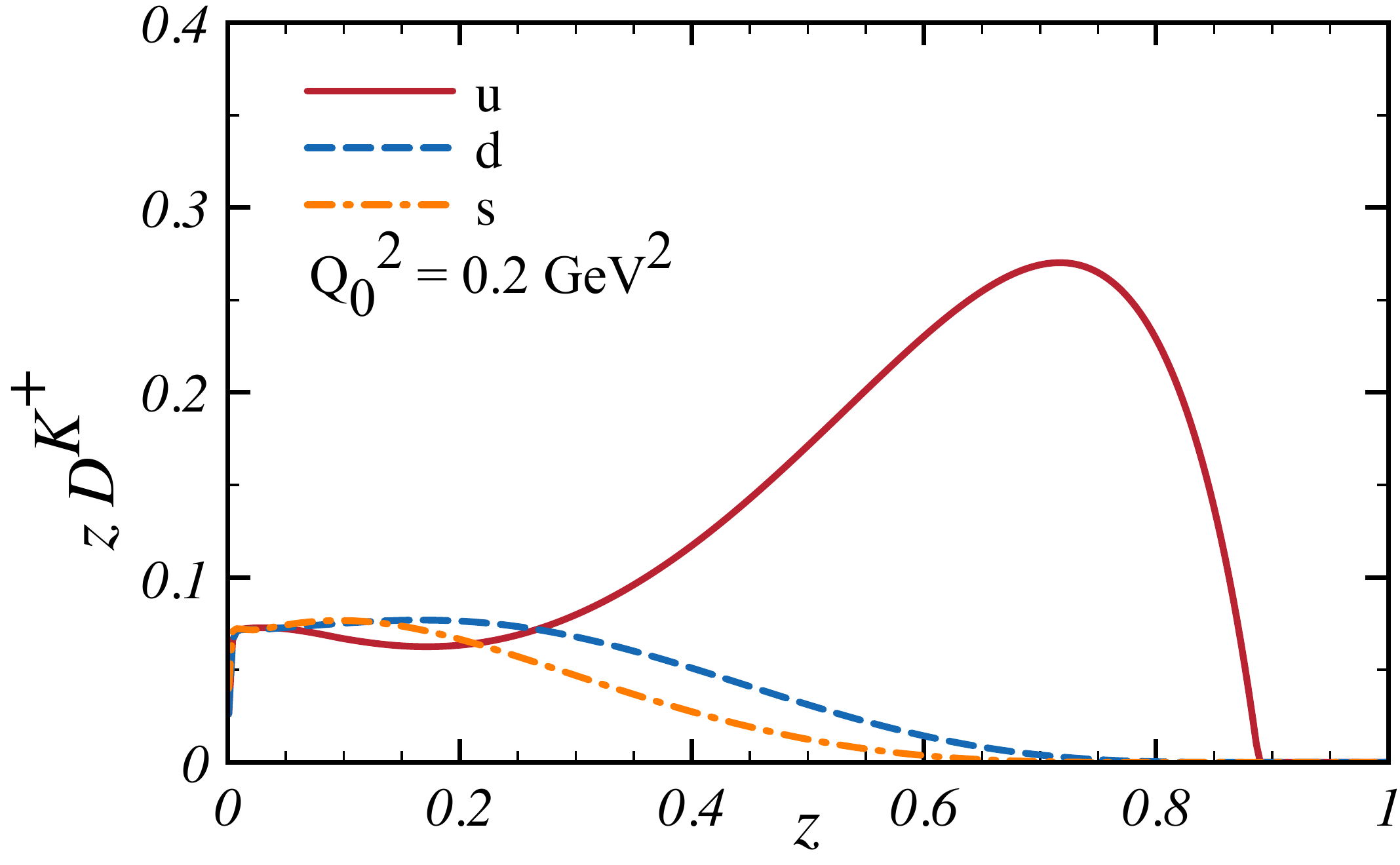}
}
\caption{NJL-Jet calculations for a) $\pi^{+}$ and b) $K^{+}$ fragmentation functions at the model scale $Q_{0}^{2}=0.2~{\rm GeV}^{2}$  as solutions of the Eq.~(\ref{EQ_JET_INT}).}
\label{PLOT_PK_FRAG}
\end{figure}

The results for the fragmentation functions of $u$, $d$ and $s$ quarks to $\pi^{+}$ and $K^{+}$ at the model scale are shown in Fig.~\ref{PLOT_PK_FRAG}. Though  the fragmentation functions of the unfavored strange and light quarks are of the same order of magnitude, we can see a notable difference between them, even at the low scale of the model. This is potentially significant observation as it has hitherto been necessary in phenomenological analyses to assume that $D_{d}^{\pi^{+}}=D_{s}^{\pi^{+}}$, etc. (e.g. see Ref.~\cite{Miyama:1995bd}) .

\begin{figure}[ptb]
\centering 
\subfigure[] {
\includegraphics[width=0.48\textwidth]{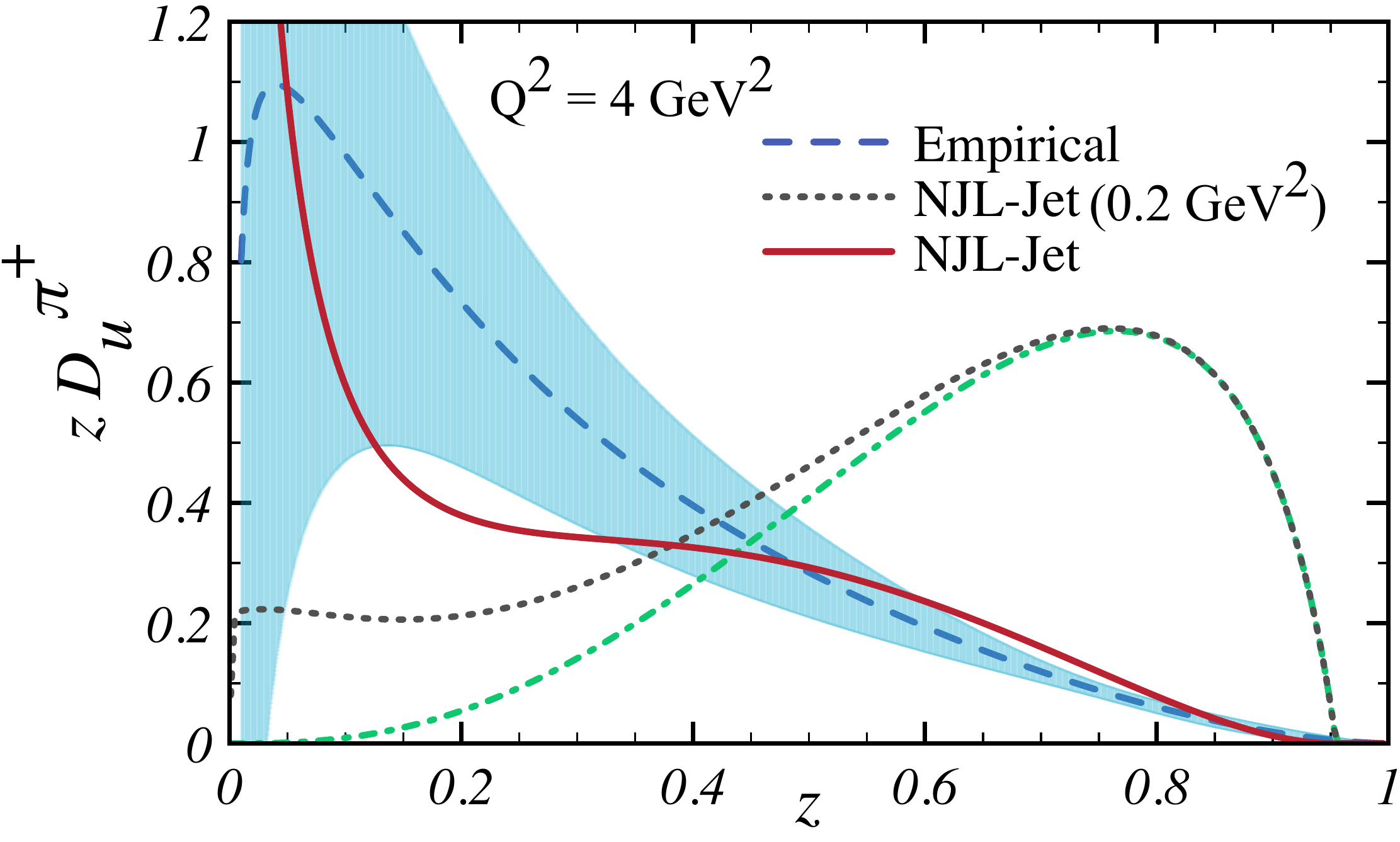}}
\hspace{0.1cm} 
\subfigure[] {
\includegraphics[width=0.48\textwidth]{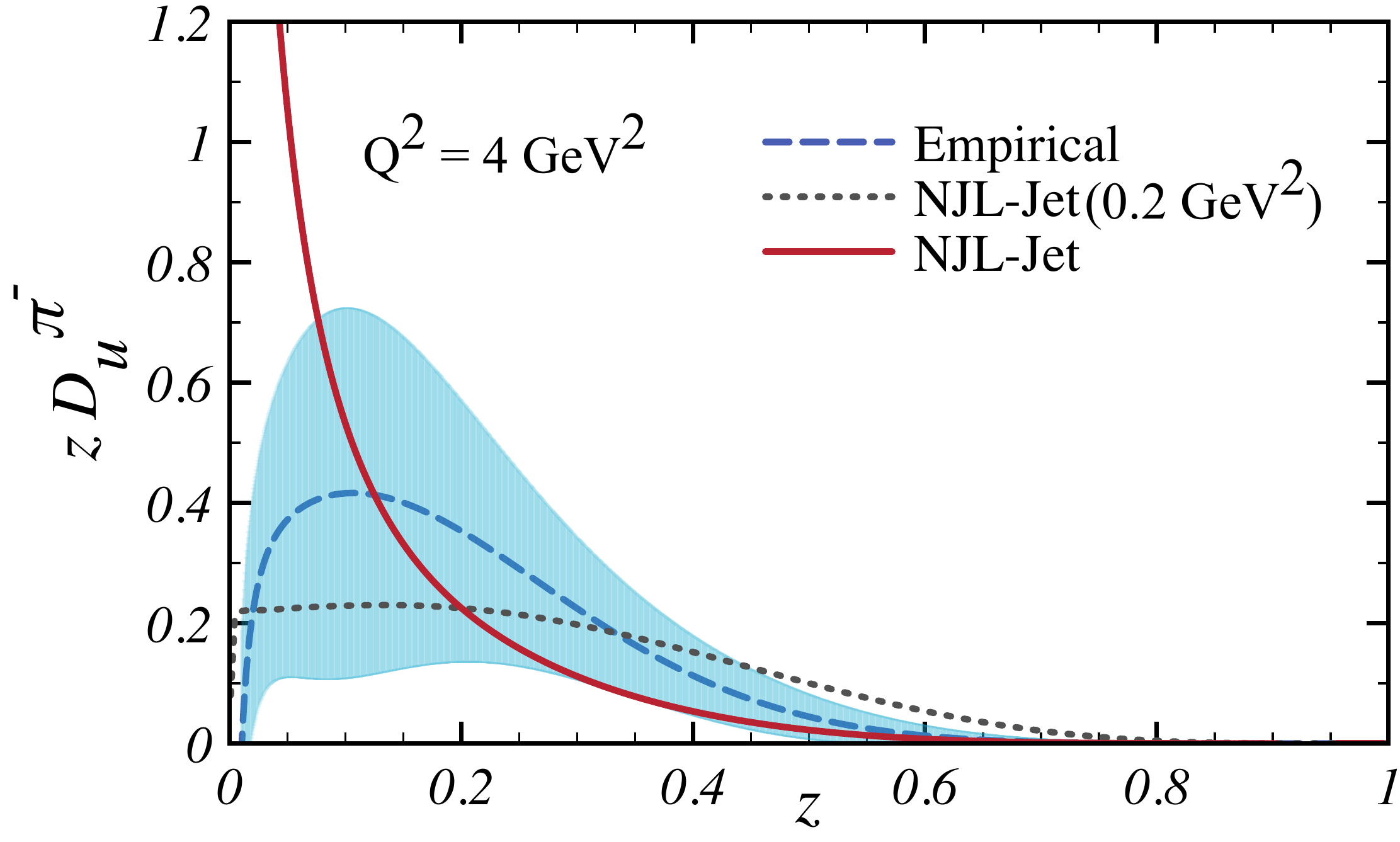}
}
\caption{Pion fragmentation functions for a) favored $zD_u^{\pi^+}(z)$ and b) unfavored  $zD_u^{\pi^-}(z)$ process . Here the dashed line with the error band represents the empirical parametrization of the Ref.~\cite{Hirai:2007cx}, the doted line is the solution for the NJL-Jet model at the low energy scale  $Q_0^2 = 0.2\ \mathrm{GeV}^2$  and the solid line represents the same solution evolved at next to leading order to the scale $Q^2 = 4\ \mathrm{GeV}^2$. The dash-dotted line is the driving term $z\ \hat{d}_u^m(z)$ in integral equation (\ref{EQ_JET_INT}). }
\label{PLOT_PION_FRAG}
\end{figure}

\begin{figure}[ptb]
\centering 
\subfigure[] {
\includegraphics[width=0.48\textwidth]{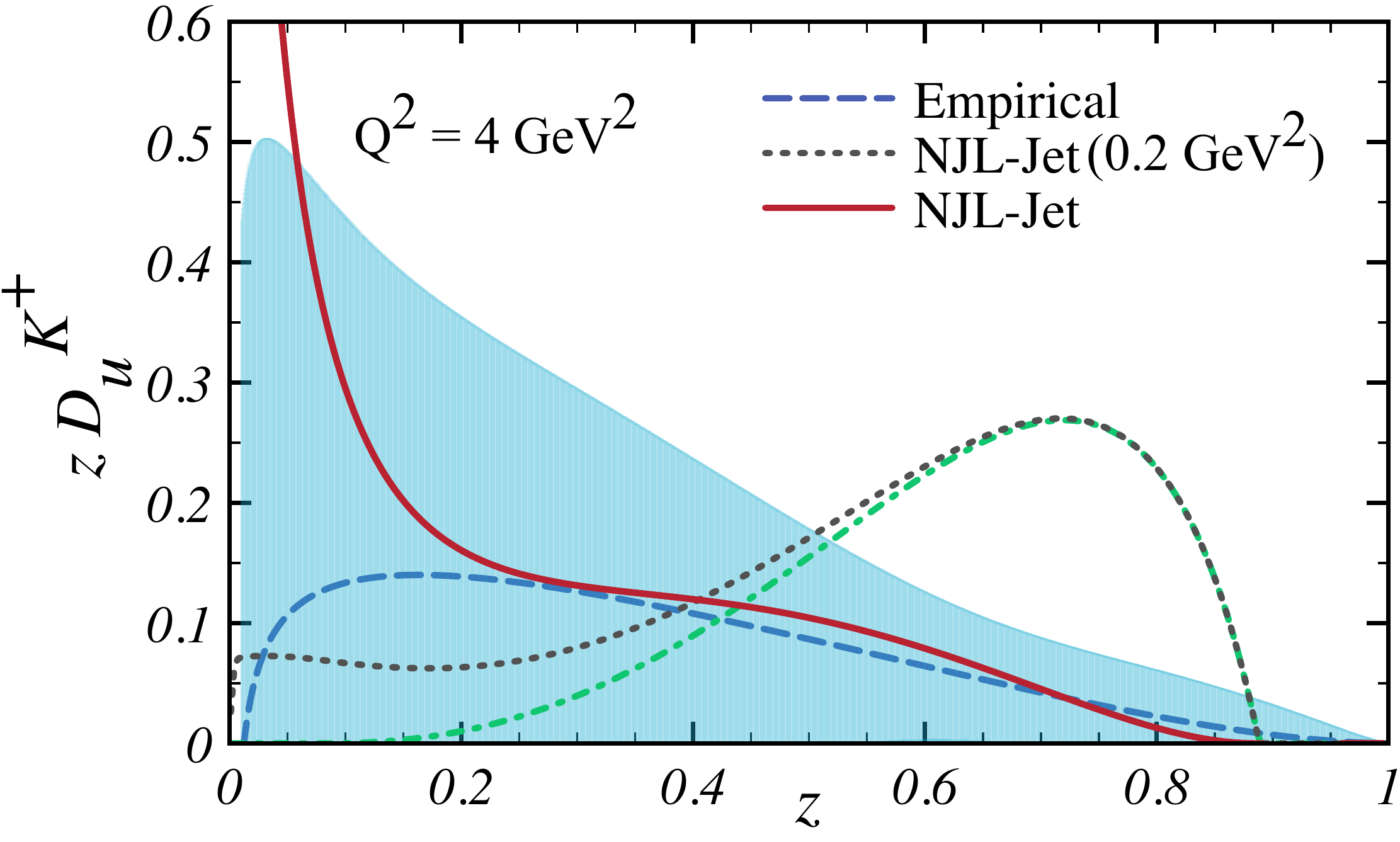}}
\hspace{0.1cm} 
\subfigure[] {
\includegraphics[width=0.48\textwidth]{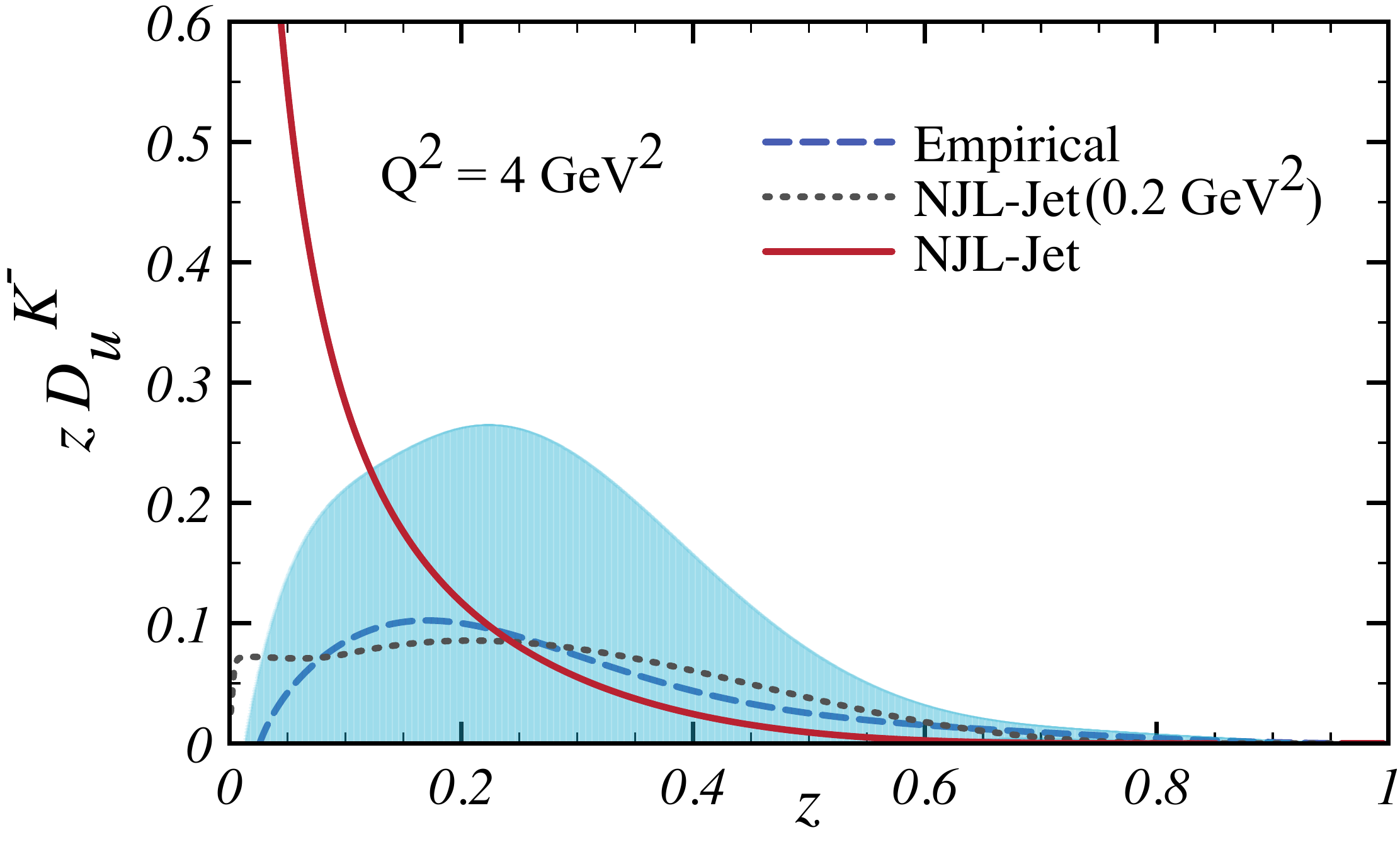}
}
\caption{Kaon fragmentation functions for a) favored $zD_u^{K^+}(z)$ and b) unfavored  $zD_u^{K^-}(z)$ process . Here the dashed line with the error band represents the empirical parametrization of the Ref.~\cite{Hirai:2007cx}, the doted line is the solution for the NJL-Jet model at the low energy scale  $Q_0^2 = 0.2\ \mathrm{GeV}^2$  and the solid line represents the same solution evolved at next to leading order to the scale $Q^2 = 4\ \mathrm{GeV}^2$. The dash-dotted line is the driving term $z\ \hat{d}_u^m(z)$ in integral equation (\ref{EQ_JET_INT}). }
\label{PLOT_KAON_FRAG}
\end{figure}

Further we present our results for the fragmentation functions $D_{u}^{\pi^{+}}$ and $D_{u}^{\pi^{-}}\equiv D_{\bar{u}}^{\pi^{+}}$ in Fig.~\ref{PLOT_PION_FRAG}. Here the DGLAP evolved curve is also compared to the empirical parametrizations of the experimental data of Ref.~\cite{Hirai:2007cx}, evolved to the same scale. We can see that the inclusion of strangeness softens the high $z$ region of $D_{u}^{\pi^{+}}$ compared to previous calculations \cite{Ito:2009zc}, thus bringing the curves closer to the phenomenological parametrizations (here we shall note that the model fragmentation functions were evolved at NLO in the current article versus the LO evolved solutions of the Ref.~\cite{Ito:2009zc} ). This is expected as the elementary fragmentation to a kaon is not negligible. In fact the fragmentation to $K^{+}$ is about half as likely as that to $\pi^{+}$ as can be seen from the plots in Fig.~\ref{PLOT_KAON_FRAG}. The plots show a reasonably good agreement with the parametrizations within the large uncertainties of the latter. 

 It is clear that for a more complete description of the quark fragmentation both vector meson and nucleon anti-nucleon channels need to be included in the calculations. This can be accomplished within the current framework.  The high $z$ region of fragmentation functions are dominated by ``few-step'' transitions where the availability of the additional fragmentation channels might have a noticeable effect. 
 
  Another limitation of the model is the assumption of the momentum scaling of the probability of hadron creation in each step of the decay chain, which is clearly only the case in the Bjorken limit. A quark with a finite momentum loses energy with each production of a hadron and finally recombines with the remnants of the antiquark jet to form the final hadron. A more accurate description of the process requires Monte-Carlo (MC) simulations of the quark fragmentation, similar to the studies in Refs.~\cite{Field:1976ve, Ritter:1979mk}, and others. MC simulations would also allow one to access the transverse momentum distribution of the produced hadrons, thus becoming relevant for the analysis of a large variety of semi-inclusive data.

 \section{Acknowledgements}
 
This work was supported by the Australian Research Council through the grant of an Australian Laureate Fellowship to A.W. Thomas and by a Subsidy for Activating Educational Institutions from the Department of Physics, Tokai University.

\appendix
\section{Flavor Factors for Distribution and Splitting Functions}

The flavor factors given in the Table~\ref{TB_FLAVOR_FACTORS} are calculated from the corresponding $SU(3)$ flavor matrices (for details see for e.g. Ref.~\cite{Klimt:1989pm}): 

\begin{table}[hpt]
\caption{Flavor Factors $C_q^m$}
\label{TB_FLAVOR_FACTORS}
\begin{center}
\resizebox{0.4\textwidth}{!}{
\begin{tabular}{|c | c|c|c|c|c|c|c|}
\hline  $C_q^m$ & $\pi^0$ & $\pi^+$ & $\pi^-$ & $K^0$ & $\overline{K}^0$ & $K^+$ & $K^-$ \\
\hline $u$ & 1 & 2 & 0 & 0 & 0 & 2 & 0 \\
\hline $d$ & 1 & 0 & 2 & 2 & 0 & 0 & 0 \\
\hline $s$ & 0 & 0 & 0 & 0 & 2 & 0 & 2\\
\hline $\overline{u}$ & 1 & 0 & 2 & 0 & 0 & 0 & 2 \\
\hline $\overline{d}$ & 1 & 2 & 0 & 0 & 2 & 0 & 0 \\
\hline $\overline{s}$ & 0 & 0 & 0 & 2 & 0 & 2 & 0 \\
\hline
\end{tabular}
}

\end{center}
\end{table}

\bibliographystyle{apsrev}
\bibliography{fragment}

\end{document}